\documentclass[aps,prf,superscriptaddress,nofootinbib,amsmath,amsfonts,amssymb,letterpaper]{revtex4-2}

\usepackage{enumerate}
\usepackage{graphicx}
\usepackage{amsmath,amssymb,amsfonts}
\usepackage{amsthm}
\usepackage{mathrsfs}
\usepackage{xcolor}
\usepackage{mathtools}
\usepackage{caption}
\usepackage{hyperref}
\hypersetup{colorlinks=true,linkcolor=blue,urlcolor=blue,citecolor=blue}

\newcommand{\Set}[1]{\mathbb{#1}}

\global\long\def\V#1{\boldsymbol{#1}} 
\global\long\def\M#1{\boldsymbol{#1}} 
\newcommand{\tp}{^\intercal}

\newcommand{\sdim}{D}
\newcommand{\node}{\mathrm{node}}
\newcommand{\edge}{\mathrm{edge}}

\global\long\def\norm#1{\left\Vert #1\right\Vert }

\global\long\def\grad{\M{\nabla}}

\begin{document}
\title{A Graph Neural Network Simulation of Dispersed Systems}
	
\author{Aref Hashemi}
\email{aref.hashemi@nd.edu}
\affiliation{Department of Applied and Computational Mathematics and Statistics, University of Notre Dame, Notre Dame, IN, United States}
\affiliation{Courant Institute, New York University, New York, NY, United States}
\author{Aliakbar Izadkhah}
\affiliation{Department of Chemical Engineering, Carnegie Mellon University, Pittsburgh, PA, United States}
    	
	
\begin{abstract}
We present a Graph Neural Network (GNN) that accurately simulates a multidisperse suspension of interacting spherical particles. Our machine learning framework is built upon the recent work of Sanchez-Gonzalez et al. ICML, PMLR, \textbf{119}, 8459--8468 (2020) on graph network simulators, and efficiently learns the intricate dynamics of the interacting particles. Nodes and edges of the GNN correspond, respectively, to the particles with their individual properties/data (e.g., radius, position, velocity) and the pairwise interactions between the particles (e.g., electrostatics, hydrodynamics). A key contribution of our work is to account for the finite dimensions of the particles and their impact on the system dynamics. We test our GNN against a representative case study of a multidisperse mixture of two-dimensional spheres sedimenting under gravity in a liquid and interacting with each other by a Lennard-Jones potential. The present GNN framework offers a fast and accurate method for the theoretical study of complex physical systems such as field-induced behavior of colloidal suspensions and ionic liquids. Our implementation of the GNN is available on GitHub at \href{https://github.com/rfjd/GNS-DispersedSystems}{github.com/rfjd/GNS-DispersedSystems}.
\end{abstract}
	
\maketitle

\section{Introduction}
Accurate and efficient simulation of suspensions and emulsions (e.g., colloidal dispersions \cite{Russel1991}) is essential for better understanding the underlying physics of these systems, as well as leveraging them for practical applications. These applications span diverse fields, from engineering to biology, including cargo-loading and targeted drug-delivery by micro/nano-robots \cite{Yang2018,cargoLoad2020,AnkurNanoscale2023}, field-induced self-assembly \cite{selfAssemblyreviewB,selfAssemblyreviewA}, and electrophoretic deposition \cite{EPD1,EPD2,EPD3}.

Such problems are governed by complex and often poorly understood physics, making it challenging to develop comprehensive mathematical models without relying on unrealistic assumptions. For instance, several long-standing open questions persist regarding the field-induced behavior of colloidal dispersions. Despite extensive research over the past two decades, theoretical and computational predictions remain `qualitatively' at odds with the experimental observations. For example, computational studies based on current theoretical models fail to predict even the direction of flow in microfluidic systems (AC electroosmosis pumps), let alone its magnitude, highlighting fundamental inaccuracies in the underlying theory \cite{Bazantreview2009,Squires2009,Garcia-Sanchez2009,Aref2020}. Similarly, available theories in electrokinetics fail to predict some of the experimentally established behaviors of charged colloids near electrodes, including the electrolyte-dependent aggregation or separation of colloids \cite{Woehl2014,Aref2018} and the extreme levitation of colloids against gravity \cite{Woehl2015,Bukosky2015,Aref2018,Bukosky2019}. To further compound matters, there is significant uncertainty regarding the validity of different modeling frameworks (e.g., continuum, Brownian Dynamics, or Molecular Dynamics), and no consensus exists among researchers on which framework is most suitable for various problems.

Moreover, even if there is a clear understanding of the underlying physics and a comprehensive mathematical model exists, one must accurately and efficiently solve the governing equations to simulate the problem and make predictions. A key challenge in this step is the computational cost, which depends on several factors, including the numerical algorithm (e.g., whether spectral or finite difference/element methods are used, the order of the finite difference/element stencils, and the choice of integration methods), the problem's complexity (e.g., whether the system involves multiphase flows, turbulence, compressible fluids, or specific types of interactions), the available computational resources, and the implementation details (e.g., programming language and use of parallel computing). Accurate simulation of multiphysics problems can take anywhere from hours to weeks or even months to perform. See a number of specific examples in water/oil emulsion \cite{CompTimeWaterOil}, turbulent multiphase flows \cite{CompTimeTurbulent}, colloidal dispersions \cite{Bukosky2019}, and fluidized beds \cite{FluidizedBedBruno,FluidizedBedMolaeia,FluidizedBedMolaeib}. For the latter, as communicated by the corresponding authors, the computation time can range from several hours \cite{FluidizedBedBruno} to weeks \cite{FluidizedBedMolaeia} and even a few months \cite{FluidizedBedMolaeib}, depending on factors such as the number of particles, simulation time, numerical algorithm, and computational resources. Note further that in some cases, the computational cost renders any realistic simulation of the problem virtually impossible. See the concluding remarks in Bukosky et al. \cite{Bukosky2019} for an example.

\vspace{1cm}
An alternative approach that circumvents these challenges is to use Machine Learning (ML) to develop \emph{data-driven} models. The incorporation of ML techniques has transformed the simulation and modeling of multiphysics systems, leading to significant improvements in computational efficiency, scalability, and generalization. These advancements can be broadly classified into two categories. The first category involves hybrid methods that combine ML with traditional Computational Fluid Dynamics (CFD) frameworks. A prominent example is the work of Kochkov et al. \cite{Kochkov2021}, where ML algorithms were integrated into traditional CFD solvers to replace components that are sensitive to coarse resolutions. This approach achieves $40$--$80$ times computational speed-ups with grids up to $10$ times coarser, while maintaining high accuracy, stability, and generalization to unseen flow conditions.

The second category consists of purely data-driven approaches. In a recent seminal work, Sanchez-Gonzalez et al. \cite{GNS-deepmind} used Graph Neural Networks (GNNs) \cite{GNN-neuralmp} to simulate the complex physics of fluid dynamics under a wide range of conditions. The corresponding ML framework, referred to as a Graph Network Simulator (GNS), treats the fluid medium (along with the immersed solid objects) as an ensemble of point particles interacting with each other. In this regard, GNS is similar to particle-based methods such as smoothed particle hydrodynamics \cite{SPH}, position based dynamics \cite{PBD}, and material point method \cite{MPM}, which also model materials as ensembles of interacting point particles. This framework has since been extended in various directions. For instance, Sergeev et al. \cite{Sergeev2024} proposed a hierarchical GNN with self-attention pooling, which further improves scalability and accuracy for complex fluid simulations. Bhattoo et al. \cite{Bhattoo2023} introduced the idea of embedding Lagrangian mechanics directly into the graph structure, enabling interpretable modeling of kinetic and potential energy alongside dissipative forces. When long-range, many-body interactions are crucial, Ma et al. \cite{Ma2022} introduced the idea of higher-order graph connectivity for hydrodynamic interactions in Stokes suspensions. Also, Martinkus et al. \cite{Martinkus2021} developed a hierarchical GNN capable of reducing particle simulation complexity to linear time and space, enabling efficient large-scale simulations of such systems. On the other hand, when extremely short-range and localized interactions are dominant, Filiatraut et al. \cite{Filiatraut2023} proposed a framework utilizing a neighborhood selection process, where each selected neighborhood is assigned its own dedicated GNN, for the study of self-assembly in microscale systems. These innovations highlight the growing impact of hybrid ML and GNN approaches across diverse scales and systems.

The main objective of the present work is to extend the GNS \cite{GNS-deepmind,GNS-pytorch} by accounting for the \emph{finite size} of the particles (droplets) in a fluid suspension (emulsion), as opposed to treating them as point particles. This extension centers on designing an innovative set of node and edge features, along with slight modifications to the network architecture, enabling the GNN to robustly learn the effects of particle size on particle-particle and particle-wall interactions. Our methodology serves as an accurate, efficient, surrogate for the often time-consuming, conventional, computational methods, such as CFD, for the study of dispersed systems. The implications of our work are vast. First, as mentioned above, the behavior of colloidal dispersions under electric fields is poorly understood, leaving a number of pressing open questions that hinder the development of electrokinetic systems for important applications in biology and engineering, such as targeted drug delivery and self-assembly. Second, our approach accommodates particles of varying sizes, allowing for the simulation of multidisperse systems. This capability is crucial because many real-world dispersed systems, such as oil-water emulsions in water purification and essential oil extraction, as well as granulation processes in fluidized beds, are inherently multidisperse. Third, the typical computation time for the simulation of such dispersed systems under practical conditions ranges from weeks to months. Our GNN framework can learn the complex physics of these systems from experimental data (or accurate simulation data if applicable) and accurately simulate their dynamics within seconds.

  We represent the immersed particles as nodes of a graph, along with their key dynamic and intrinsic properties such as position, velocity, and radius. The interaction between the particles is modeled as edges of the graph; i.e., the pair-interaction between the particles is represented as a message-passing process between the corresponding nodes of the graph. We evaluate the accuracy of our GNN through a representative case-study of a multidisperse suspension of particles sedimenting under gravity in a box of fluid. A comparison between the ground-truth data and the predictions of the trained GNN shows excellent agreement. The trained GNN accurately and efficiently captures the most intricate characteristics of the system behavior, including individual particle trajectories, clusters, collisions, and boundaries. Our implementation of the GNN is publicly available on GitHub at \href{https://github.com/rfjd/GNS-DispersedSystems}{github.com/rfjd/GNS-DispersedSystems}.

The remainder of this article is structured as follows. In \autoref{sec:GNNFramework}, we describe the Graph Neural Network (GNN) framework that we have developed for this study, including the network architecture and the specific node and edge features used to model the system dynamics and particle interactions. \autoref{sec:data} presents a case study of a multidisperse suspension sedimenting under gravity in a box of fluid, where we use a numerical solution to generate the sample training data. In \autoref{sec:res}, we evaluate the performance of the GNN in accurately predicting the dynamics of the system and compare its results to the ground-truth data. Finally, in \autoref{sec:conc}, we discuss the implications of our findings, address potential limitations, and provide concluding remarks on the utility and future applications of our approach.

\begin{figure}[t]
  \centering
  \begin{minipage}{0.7\textwidth} 
    \centering
    \setlength{\belowcaptionskip}{-10pt}
    \captionsetup{justification=raggedright, singlelinecheck=false}
    \includegraphics{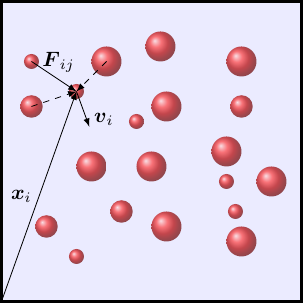}
    \captionof{figure}{Schematic diagram of the problem: a multidisperse suspension of interacting spherical particles in a rectangular box of liquid. A target particle $i$, with current position $\V{x}_i$ and velocity $\V{v}_i$, moves and interacts with its neighbors.}
    \label{fig:schematic}
  \end{minipage}
\end{figure}

\section{GNN Framework}\label{sec:GNNFramework}
\subsection{Multidisperse System of Interacting Spheres}
We consider $N$ spherical particles (or droplets) with finite radii $a_i,\;i=1,2,\cdots,N$, interacting and moving in a box of liquid (see \autoref{fig:schematic}). Depending on the size and charge of the particles, and the type of the liquid, various interactions such as electrostatics, hydrodynamics, steric, and van der Waals forces can be present. Also, the particles experience a drag force from the liquid medium, and may move due to spatially and temporally uniform or nonuniform external forces, such as those from a gravitational field or an imposed electric field.

The training data contains trajectories of all particles with a time-step $\Delta t$. Such training data can be obtained from experiments or computer simulations. The objective is to train a GNN to learn the dynamics of such a system and accurately predict the trajectories of all particles. In the case of computer simulations, the idea is that the trained GNN (i.e., our data-driven model) is much faster than the physics-based simulator. If there are no accurate computer simulations for a particularly complex problem (due to a knowledge gap in theory or high computational cost), experimental data can be used. As shown in \autoref{fig:schematic}, we denote the position and velocity of particle $i$ with $\V{x}_i$ and $\V{v}_i$, respectively. Furthermore, $\V{F}_{ij}$ is the force on particle $i$ due to the neighboring particle $j$. We assume that we do not have any information about the nature of this force; instead, we see the outcome of the interaction through the trajectories of the particles. Following Sanchez-Gonzalez et al. \cite{GNS-deepmind}, we use a first-order finite difference scheme to extract particle velocities $\V{v}_i$ from positions $\V{x}_i$.

For a physics-based model, predicting the next position of the particles requires two \emph{initial} conditions, typically the current positions and velocities. Using a first-order finite difference scheme to compute velocities from positions, this translates to a sequence of two positions of all particles (i.e., $\{\V{x}(t-2\Delta t),\V{x}(t-\Delta t)\}\to \V{x}(t)$). Indeed, this information is sufficient for the GNN to predict the particle trajectories as well. However, using longer sequences has been shown to significantly improve the prediction accuracy \cite{GNS-deepmind}. Unless stated otherwise, for the results presented here, we have used a sequence of $C=6$ most recent positions to predict the next position. Consequently, a single full-length computational or experimental positional dataset with $M$ time steps results in $(M-6)$ training trajectories (or `examples').

\subsection{Node \& Edge Features}
We closely follow the work of Sanchez-Gonzalez et al. \cite{GNS-deepmind} to assign the node and edge features with some key differences that allows an accurate capture of particle size, as well as particle-particle and particle-wall interactions.

\begin{figure}[t]
  \centering
  \setlength{\belowcaptionskip}{-10pt}
  \includegraphics{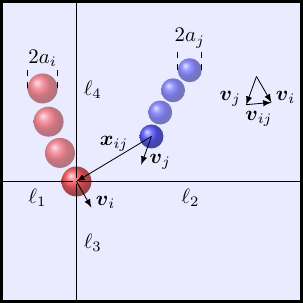}
  \caption{Node and edge features of the GNN.}
  \label{fig:features}
\end{figure}

\textbf{Node Features.} The node features consist of (i) the last $C-1$ velocities of each particle, $\V{v}_i$ ($(C-1)\times\sdim$ values, where $\sdim$ denotes the spatial dimension of the system), (ii) the distances to the boundaries ($\ell_i$'s in \autoref{fig:features}; $2\times\sdim$ values), and (iii) the particle radius $a_i$ ($1$ value). These features are selected to characterize each particle's location relative to the boundaries, and the evolution of its velocity over the last $C-1$ steps. To facilitate learning, we normalize the velocities and distances corresponding to each particle by its radius. We acknowledge that a more physically appropriate normalization would be a constant length scale for all particles. But our scaling approach, combined with the method for defining the edge features (described in the next paragraph), proves effective in enabling the GNN to accurately learn critical system characteristics, such as clustering. 

\textbf{Edge Features.} The edge features for a pair of interacting particles $i$ and $j$ include (i) the relative displacement $\V{x}_{ij}=\V{x}_i-\V{x}_j$ and absolute distance $\norm{\V{x}_{ij}}$ (see \autoref{fig:features}; $\sdim+1$ values), (ii) the last $C-1$ relative velocities $\V{v}_i-\V{v}_j$ ($(C-1)\times\sdim$ values), and (iii) the sum of the particle radii $a_i+a_j$ ($1$ value). We normalize all velocities and distances by $a_i+a_j$. These features capture the relative positions of particles and their motion with respect to each other over the last $C-1$ steps. This combination of features ensures that the GNN incorporates the key factors influencing pairwise interactions, such as size asymmetry, proximity, and relative dynamics.

\subsection{GNN Architecture}
\begin{figure}[h]
  \centering
  \setlength{\belowcaptionskip}{-10pt}
  \includegraphics{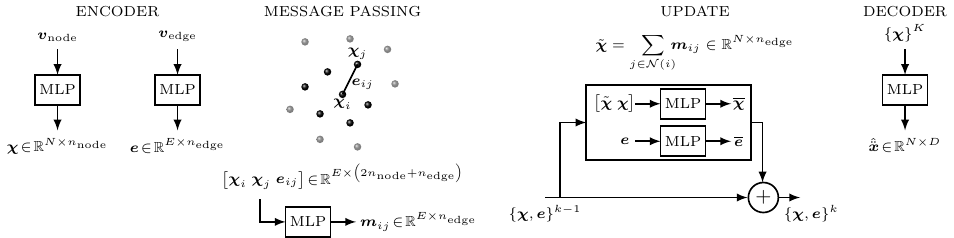}
  \caption{An outline of our network architecture, including the encoding, message passing, and decoding steps.}
  \label{fig:architecture}
\end{figure}

Our GNN architecture is largely similar to that developed by Sanchez-Gonzalez et al. \cite{GNS-deepmind}, and in particular its PyTorch implementation by Kumar and Vantassel \cite{GNS-pytorch}. An overview of the network architecture is provided in \autoref{fig:architecture}. We store the node and edge features as one-dimensional column vectors $\V{v}_{\node}\in\Set{R}^{(C+1)\sdim+1}$ and $\V{v}_{\edge}\in\Set{R}^{C\sdim+2}$, respectively. These vectors are encoded by two Multilayer Perceptrons (MLPs) into their corresponding embeddings $\V{\chi}\in\Set{R}^{N\times n_{\node}}$ and $\V{e}\in\Set{R}^{E\times n_{\edge}}$, respectively. Here, symbols stand for number of nodes, $N$; number of edges, $E$; node embedding size, $n_{\node}$; and edge embedding size, $n_{\edge}$. The node and edge embedding sizes are set to $128$ and kept the same throughout the network. We then perform a total of $K$ message passing steps as follows. In each step of the message passing from a neighbor node $j$ to a target node $i$, we concatenate $\V{\chi}_i$, $\V{\chi}_j$, and $\V{e}_{ij}$ and pass it through an MLP. Then we aggregate all messages from the neighbors of the target node $i$ (i.e, $j\in\mathcal{N}(i)$) to find an overall message $\tilde{\V{\chi}}\in\Set{R}^{N\times n_{\edge}}$. Note that the aggregation of all messages is inspired by the physics of the problem, where each neighboring particle exerts a force on the target particle, and the total force is the sum of these interaction forces. Following prior work, we finally apply a residual connection to update the node and edge embeddings. The final node embedding after $K$ message passing steps, $\{\V{\chi}\}^{K}$, is then passed through a final MLP to predict the particle accelerations $\hat{\ddot{\V{x}}}\in\Set{R}^{N\times\sdim}$.

During training, the objective of the model is to predict the acceleration of the particles, given a sequence of the last $C$ positions. The loss function to minimize during training is
\begin{equation}
  \mathcal{L}=\frac{1}{N}\norm{\ddot{\V{x}}-\hat{\ddot{\V{x}}}}_2^2,
  \label{eq:loss}
\end{equation}
where $\ddot{\V{x}}$ and $\hat{\ddot{\V{x}}}$ are the true and predicted accelerations, and $N$ is the number of particles in the example. Following Sanchez-Gonzalez et al. \cite{GNS-deepmind}, we corrupt the sequence of particle positions with a random walk noise, and use finite difference to compute the accelerations from true positions. For rollout predictions during inference, we use the same finite difference stencil to compute the particle positions from the predicted accelerations.

It is important to note that similar to the prior work by Sanchez-Gonzalez et al. \cite{GNS-deepmind}, our GNN implementation assumes a time-step of $\Delta t=1$. Consequently, particle accelerations at any step $p$ are computed as $\ddot{\V{x}}(p)=\V{x}(p)-2\V{x}(p-1)+\V{x}(p-2)$ (second-order backward finite difference scheme with a unit time-step). Note that this does not cause any issues with the training. When interpreting the network output during inference, the true time-step $\Delta t$ can be incorporated to make predictions in correct units and dimensions.

We have not performed an extensive optimization to fine-tune the network parameters such as the number of message passing steps, number of hidden layers and neurons, etc. However, we found the parameters that work sufficiently well for our objectives. Please see the default parameter values in \href{https://github.com/rfjd/GNS-DispersedSystems}{github.com/rfjd/GNS-DispersedSystems}.

\section{Sample Dataset Generation}\label{sec:data}
In this section, we use a toy model to generate synthetic `ground truth' data for training and testing our GNN. The model represents a multidisperse suspension of spherical disks interacting and sedimenting under gravity within a two-dimensional box $x,y\in[-L,L]$. Computer simulations are performed to generate the corresponding training and test datasets. It is important to note that this case study is not intended to provide a physically accurate representation of the system. However, it has the key characteristics to challenge our GNN model, such as particles of different sizes, intricate interactions between the particles, and physical boundaries. Alternatively, one could generate the training and test datasets from actual experimental data or by using a black-box multiphysics simulator like Ansys Fluent.

We describe the motion of the particles using the Newton's second law:
\begin{subequations}
\begin{align}
  \pi a_i^2s\rho\frac{d\V{v}_i}{dt}&=-k a_i\V{v}_i+\V{F}_i^{\text{external}}(\V{x})+\V{F}_i^{\text{interaction}}(\V{x}),\\
  \frac{d\V{x}_i}{dt}&=\V{v}_i.
\end{align}
\label{eq:equation-motion}%
\end{subequations}
Note that we consider the deterministic motion of the particles, with no stochastic processes such as thermal fluctuations involved. Here the symbols stand for the velocity of the particles, $\V{v}=\begin{bmatrix}\V{v}_1,\V{v}_2,\dots\end{bmatrix}^{\tp}\in\Set{R}^{N\times 2}$; position of the particles, $\V{x}=\begin{bmatrix}\V{x}_1,\V{x}_2,\dots\end{bmatrix}^{\tp}\in\Set{R}^{N\times 2}$; vectors of, respectively, external, $\V{F}^{\text{external}}(\V{x})\in\Set{R}^{N\times 2}$, and interaction, $\V{F}^{\text{interaction}}(\V{x})\in\Set{R}^{N\times 2}$, forces; thickness, $s$, and density, $\rho$, of the particles; and time, $t$. For this `toy' problem, we assume that the drag force on each particle is proportional to the particle's velocity and radius with a constant of proportionality $k$.

For simplicity, we suppose that the external forces are constant and due to gravity only,
\begin{equation}
  \V{F}_i^{\text{external}}=\begin{bmatrix}0 & -\pi a_i^2s(\rho-\rho_f)g\end{bmatrix},
  \label{eq:external-force}
\end{equation}
where $\rho_f$ is the density of the medium, and $g$ is the gravitational acceleration. We further consider a modified, truncated, Lennard-Jones (LJ) potential to model the interactions between particles. Specifically, for particles $i$ and $j$,
\begin{equation}
  U(r_{ij})=
  \begin{cases}
    U_{\text{LJ}}(r_m)+\frac{dU_{\mathrm{LJ}}(r_m)}{dr_{ij}}(r_{ij}-r_m) & r_{ij}\leq r_m,\\
    U_{\text{LJ}}(r_{ij}) & r_m<r_{ij},
  \end{cases}
  \label{eq:LJpot}
\end{equation}
with
\begin{equation}
  U_{\text{LJ}}(r_{ij})=4U_0\left(\left(\frac{\sigma_{ij}}{r_{ij}}\right)^{2p}-\left(\frac{\sigma_{ij}}{r_{ij}}\right)^p\right).
  \label{eq:LJpot}
\end{equation}
The symbols stand for the particle-particle distance, $r_{ij}$; cutoff distance, $r_m$; and depth of the potential well, $U_0$. The lower cutoff distance $r_m<\sigma_{ij}$ determines the maximum repulsive force for a given set of parameters. We let $\sigma_{ij}=a_i+a_j$. The LJ force on particle $i$ due to particle $j$ is
\begin{align}
  \V{F}^{\text{interaction}}_{ij}(\V{x})&=-\grad U(r_{ij})=\nonumber\\
  &\begin{cases}
    -\frac{1}{r_{ij}}\frac{dU_{\mathrm{LJ}}(r_m)}{dr_{ij}}\V{x}_{ij} & r_{ij}\leq r_m,\\
    -\frac{1}{r_{ij}}\frac{dU_{\mathrm{LJ}}(r_{ij})}{dr_{ij}}\V{x}_{ij} & r_m<r_{ij},
  \end{cases}
  \label{eq:LJforce}
\end{align}
with $\V{x}_{ij}=\begin{bmatrix}x_i-x_j & y_i-y_j\end{bmatrix}$, $r_{ij}=\sqrt{x_{ij}^2+y_{ij}^2}$, and
\begin{equation}
  \frac{dU_{\mathrm{LJ}}(r_{ij})}{dr_{ij}}=-\frac{4pU_0}{r_{ij}}\left(2\left(\frac{\sigma_{ij}}{r_{ij}}\right)^{2p}-\left(\frac{\sigma_{ij}}{r_{ij}}\right)^p\right)
  \label{Eq:LJderivative}
\end{equation}
Accordingly, the total LJ force experienced by particle $i$ is $\sum_{j\neq i}\V{F}^{\text{interaction}}_{ij}(\V{x})$.

We could also consider other types of interactions, such as hydrodynamics, or electrostatics for systems involving charged particles or ions. However, the Lennard-Jones interactions used in this toy model provide sufficient complexity in terms of stiffness and spatial dependence to serve the main goal of rigorously testing the GNN. The primary focus here is not to develop a highly accurate physical model, but to generate suitable datasets for the GNN's evaluation. In this context, the GNN will ultimately focus on learning from the data, irrespective of the specific physical model used to generate it. Nonetheless, in cases where the nature of interactions significantly differs, it may be necessary to adjust the network architecture or features to ensure effective learning.

To complete the description of the toy model, we impose solid-wall (i.e., no overlap upon collision) boundary conditions with an energy loss ratio $\gamma\in(0,1]$; i.e., upon each collision with a wall, the kinetic energy of the colliding particle is reduced to $\gamma$ times its pre-collision value. For simplicity we suppose that this energy loss affects both velocity components equally, so that the post-collision velocity is scaled by $\sqrt{\gamma}$. Furthermore, the direction of the particle motion after collision is determined by assuming a mirror reflection from the wall. Denoting the velocity vectors before and after collision as $\V{v}$ and $\V{v}'$, respectively, this reflection can be expressed as  
\begin{equation}
  \V{v}'=\sqrt{\gamma}\left[\V{v}-2\left(\V{v}\cdot\hat{\V{n}}\right)\hat{\V{n}}\right],
\end{equation}
where $\hat{\V{n}}$ is the normal unit vector pointing into the domain.

To generate particle trajectories, we use the Euler method to numerically integrate the equation of motion \eqref{eq:equation-motion}. While more accurate numerical schemes could be employed, the primary objective here is to create datasets for training and testing our GNN, rather than developing a comprehensive, highly accurate, \emph{physics-based} model or simulator.

For the simulations, we use the following parameter values: $s=2\;\mathrm{mm}$, $\rho=2300\;\mathrm{kg\cdot m^{-3}}$, $\rho_f=1000\;\mathrm{kg\cdot m^{-3}}$, $g=9.81\;\mathrm{m\cdot s^{-2}}$, $L=10\;\mathrm{cm}$, $k=1.25\;\mathrm{kg\cdot s^{-1}\cdot m^{-1}}$, and $\gamma=0.8$. We also use $\Delta t=10^{-4}\;\mathrm{s}$, but take a snapshot of the particle positions every $15$ time steps. For each simulation, a few hundred particles with randomly selected radii from a pre-specified list were placed in the upper half of the box at random positions and with random initial two-dimensional velocities. We store the positional data and particle radii scaled by the box size $L$. For further details, please refer to the corresponding code script in the repository.

\vspace{-0.25cm}
\section{Results \& Discussion}\label{sec:res}
We performed a total of $16$ full-length simulations, each generating approximately $500$ snapshots. With $C=6$, this yields $16\times(500-6)=7904$ training examples. The model was trained for $2$ million steps with a batch size of $2$ examples, equivalent to approximately $500$ epochs.

Recall that the loss function \eqref{eq:loss} compares the true and predicted accelerations of the particles, given a sequence of last $C$ positions. Our ultimate goal, however, is to make accurate rollout predictions, i.e., full trajectories of all particles over time. To quantify the error in our rollout predictions, we introduce a rollout loss, $\mathcal{L}_{\text{rollout}}$, defined as the mean squared error between the true and predicted \emph{positions} (normalized by the box size $L$), across all time steps, particles, and spatial dimensions in a full-length \emph{test} simulation.

\autoref{fig:gt-gnn-test} compares the ground-truth and the predictions of our trained GNN for a test simulation (see Supplementary Video 1 for an animation). The trained GNN accurately predicts the particle positions over long trajectories and captures important spatial structures, such as clusters and relative placements of particles. Furthermore, it learns the intricate nature of particle-particle and particle-wall interactions and collisions. Notably, the particles never cross the box boundaries, even though no mask or penalty is applied to enforce this behavior. The model's accuracy slightly drops as the particles get crowded at the bottom of the box. We argue, however, that more training data will resolve this issue. The model simply has not seen enough trajectories in such relatively more complex configurations. The rollout loss for six randomly generated test cases had a mean value of $9.5\times 10^{-4}$ with a standard deviation of $2.2\times 10^{-4}$. For the results shown in \autoref{fig:gt-gnn-test}, $\mathcal{L}_{\text{rollout}}=6.6\times 10^{-4}$. To provide insight into the physical significance of this loss, for $L=10\;\mathrm{cm}$, it translates to an average discrepancy of $0.066\;\mathrm{mm}$ between the ground-truth and predicted trajectories. For an average particle diameter of $6\;\mathrm{mm}$, this results in a mere $\approx1\%$ error relative to the particle size. Also, visually, it is quite challenging to distinguish the predicted trajectories from the ground truth, further underscoring the accuracy of the model.

We emphasize that the particle size could be treated as a categorical property, where distinct particle sizes (categories) are considered during training, and the particle radius would act as a particle type. However, we were interested in treating the particle radius as a continuous quantity, and in developing a model that predicts the dynamics of a system with any particle size. To demonstrate this point, we tested our trained model on a system with particle sizes not included in the training data. \autoref{fig:gt-gnn-test-extra1} and Supplementary Video 2 show the corresponding results. Remarkably, the GNN accurately predicts the particle trajectories despite not having seen these particle sizes during training. For $12$ test cases involving only particle sizes not present during training, the average rollout loss was $9.4\times 10^{-4}$ with a standard deviation of $2\times 10^{-4}$. This performance is noteworthy, as the GNN generates predictions for previously unseen particle sizes with a loss comparable to the predictions for particle sizes seen during training. We should note, however, that the model's accuracy diminishes for systems involving particles with radii far outside the radius range used in training. Similarly, one could use training data that include different fluids, particle materials, particle-fluid interfacial properties, etc. to develop a comprehensive model that captures the impact of various physical properties on the system.
  
\begin{minipage}[t]{0.47\textwidth}
  \centering
  \setlength{\belowcaptionskip}{-10pt}
  \includegraphics{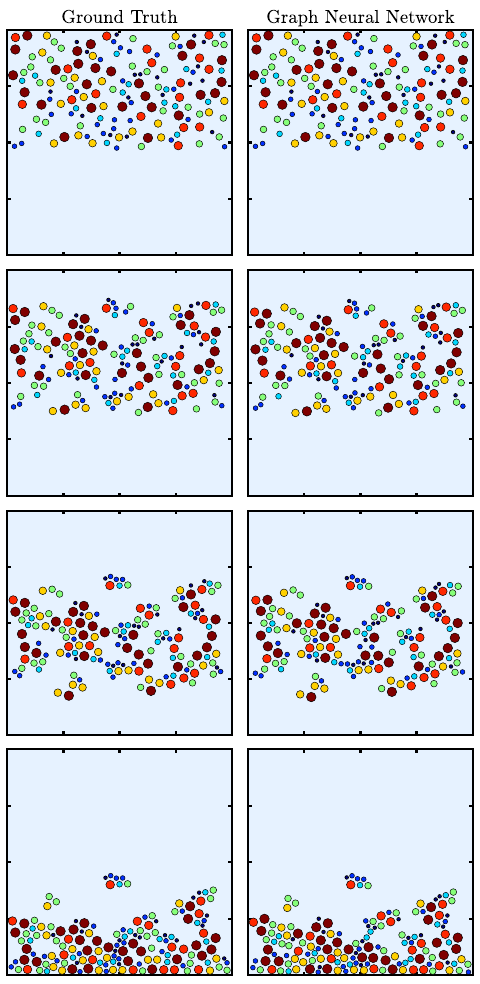}
  \captionof{figure}{Comparison of the ground-truth (left panel) and the GNN prediction (right panel) for the dynamics of a multidisperse suspension of interacting particles sedimenting under gravity in a square box filled with liquid (a \emph{test} simulation not used for training). $\mathcal{L}_{\text{rollout}}=6.6\times 10^{-4}$.}
  \label{fig:gt-gnn-test}
\end{minipage}\hfill
\begin{minipage}[t]{0.47\textwidth}
  \centering
  \setlength{\belowcaptionskip}{-10pt}
  \includegraphics{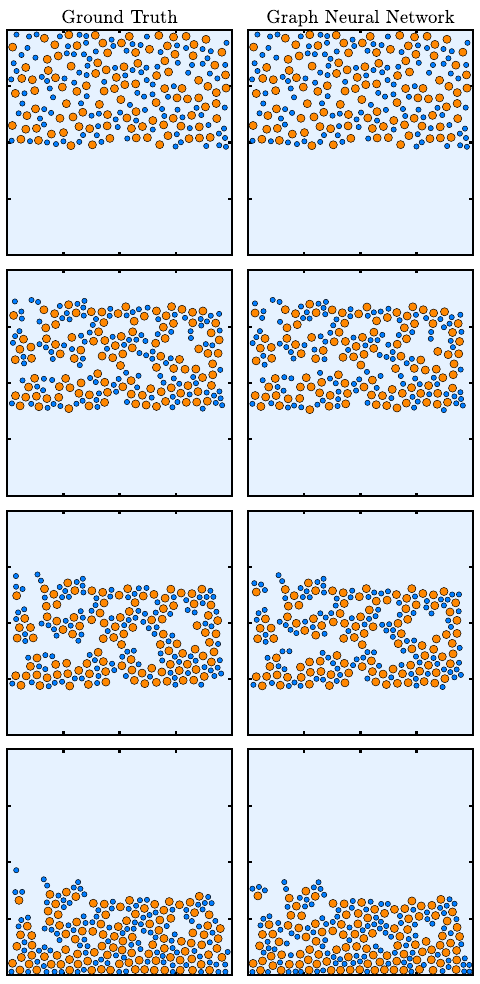}
  \captionof{figure}{Comparison of the ground truth (left panel) and the GNN prediction (right panel) for a system with particle sizes not present during training. $\mathcal{L}_{\text{rollout}}=6.5\times 10^{-4}$.}
  \label{fig:gt-gnn-test-extra1}
\end{minipage}

\section{Conclusions}\label{sec:conc}
The work presented here introduces a new ML framework, a graph neural network (GNN), for a data-driven dynamic simulation of multidisperse suspensions and emulsions in liquids. Our GNN accurately captures the most subtle characteristics of such complex systems, including clustering, as well as particle-particle and particle-wall interactions. This work opens a new avenue of research in using GNNs to simulate dispersed systems. 

We briefly discuss a few future works below.
\begin{enumerate}[(i)]
\item \textbf{Inclusion of Rotation.} An exciting future work is to account for the inevitable rotation of the particles in two and three dimensions. In this scenario, the orientation and angular velocity of the particles contribute to the node and edge features, and the GNN predicts both the translational and rotational accelerations of the particles during training. Since this problem is inherently more complex, it would put our machine learning framework into a more rigorous test. The training data can be obtained either through experiments and high-speed-imaging, or using simulation software such as Ansys Fluent. However, the former, albeit more appealing to us, faces some challenges in a three-dimensional system (e.g., particles masking each other).
\end{enumerate}

\begin{enumerate}[(i)]\setcounter{enumi}{1}
\item \textbf{Asymmetric Particles.} Our GNN can be extended to systems with non-spherical particles. One could represent particles with multiple geometrical characteristics, rather than just a single radius as we used here for spheres (e.g., length and thickness for a rod-like particle). Additionally, the relative orientation of the particles plays an important role in the particle-particle and particle-wall interactions and collisions. While we acknowledge that this is a significantly more complex problem, we believe it is a feasible future project.
\item \textbf{Colloidal Dispersions.} Another avenue for future work is the application of the proposed GNN framework to experimental colloidal dispersions. Recent advances, such as those demonstrated by ActiveNet \cite{ActiveNet}, highlight the potential of GNNs to uncover active and two-body forces in colloidal systems by learning directly from experimental trajectories. Extending our approach to incorporate these techniques could enable the study of forces and dynamics in a wider range of active and passive colloidal systems, including those under external fields or exhibiting many-body interactions.
\end{enumerate}

\section*{Acknowledgments}
We thank Dr. A. Molaei Dehkordi for insightful discussions on the time complexity of computer simulations of fluidized beds.

\section*{Statements and Declarations}
\noindent{\textbf{Competing Interests.}} There are no conﬂicts to declare.\\
\noindent{\textbf{Data Availability.}} All of the codes along with the training and test data used to write this paper can be found on GitHub at \href{https://github.com/rfjd/GNS-DispersedSystems}{github.com/rfjd/GNS-DispersedSystems}.


\begin{thebibliography}{36}%
\makeatletter
\providecommand \@ifxundefined [1]{%
 \@ifx{#1\undefined}
}%
\providecommand \@ifnum [1]{%
 \ifnum #1\expandafter \@firstoftwo
 \else \expandafter \@secondoftwo
 \fi
}%
\providecommand \@ifx [1]{%
 \ifx #1\expandafter \@firstoftwo
 \else \expandafter \@secondoftwo
 \fi
}%
\providecommand \natexlab [1]{#1}%
\providecommand \enquote  [1]{``#1''}%
\providecommand \bibnamefont  [1]{#1}%
\providecommand \bibfnamefont [1]{#1}%
\providecommand \citenamefont [1]{#1}%
\providecommand \href@noop [0]{\@secondoftwo}%
\providecommand \href [0]{\begingroup \@sanitize@url \@href}%
\providecommand \@href[1]{\@@startlink{#1}\@@href}%
\providecommand \@@href[1]{\endgroup#1\@@endlink}%
\providecommand \@sanitize@url [0]{\catcode `\\12\catcode `\$12\catcode
  `\&12\catcode `\#12\catcode `\^12\catcode `\_12\catcode `\%12\relax}%
\providecommand \@@startlink[1]{}%
\providecommand \@@endlink[0]{}%
\providecommand \url  [0]{\begingroup\@sanitize@url \@url }%
\providecommand \@url [1]{\endgroup\@href {#1}{\urlprefix }}%
\providecommand \urlprefix  [0]{URL }%
\providecommand \Eprint [0]{\href }%
\providecommand \doibase [0]{http://dx.doi.org/}%
\providecommand \selectlanguage [0]{\@gobble}%
\providecommand \bibinfo  [0]{\@secondoftwo}%
\providecommand \bibfield  [0]{\@secondoftwo}%
\providecommand \translation [1]{[#1]}%
\providecommand \BibitemOpen [0]{}%
\providecommand \bibitemStop [0]{}%
\providecommand \bibitemNoStop [0]{.\EOS\space}%
\providecommand \EOS [0]{\spacefactor3000\relax}%
\providecommand \BibitemShut  [1]{\csname bibitem#1\endcsname}%
\let\auto@bib@innerbib\@empty
\bibitem [{\citenamefont {Russel}\ \emph {et~al.}(1991)\citenamefont {Russel},
  \citenamefont {Saville},\ and\ \citenamefont {Schowalter}}]{Russel1991}%
  \BibitemOpen
  \bibfield  {author} {\bibinfo {author} {\bibfnamefont {W.~B.}\ \bibnamefont
  {Russel}}, \bibinfo {author} {\bibfnamefont {D.~A.}\ \bibnamefont {Saville}},
  \ and\ \bibinfo {author} {\bibfnamefont {W.~R.}\ \bibnamefont {Schowalter}},\
  }\href@noop {} {\emph {\bibinfo {title} {Colloidal Dispersions}}},\ \bibinfo
  {edition} {1st}\ ed.\ (\bibinfo  {publisher} {Cambridge University Press},\
  \bibinfo {address} {Cambridge, UK},\ \bibinfo {year} {1991})\BibitemShut
  {NoStop}%
\bibitem [{\citenamefont {Yang}\ and\ \citenamefont {Wu}(2018)}]{Yang2018}%
  \BibitemOpen
  \bibfield  {author} {\bibinfo {author} {\bibfnamefont {X.}~\bibnamefont
  {Yang}}\ and\ \bibinfo {author} {\bibfnamefont {N.}~\bibnamefont {Wu}},\
  }\href {\doibase https://doi.org/10.1021/acs.langmuir.7b02793} {\bibfield
  {journal} {\bibinfo  {journal} {Langmuir}\ }\textbf {\bibinfo {volume}
  {34}},\ \bibinfo {pages} {952} (\bibinfo {year} {2018})}\BibitemShut
  {NoStop}%
\bibitem [{\citenamefont {Huo}\ \emph {et~al.}(2020)\citenamefont {Huo},
  \citenamefont {Wu}, \citenamefont {Boymelgreen},\ and\ \citenamefont
  {Yossifon}}]{cargoLoad2020}%
  \BibitemOpen
  \bibfield  {author} {\bibinfo {author} {\bibfnamefont {X.}~\bibnamefont
  {Huo}}, \bibinfo {author} {\bibfnamefont {Y.}~\bibnamefont {Wu}}, \bibinfo
  {author} {\bibfnamefont {A.}~\bibnamefont {Boymelgreen}}, \ and\ \bibinfo
  {author} {\bibfnamefont {G.}~\bibnamefont {Yossifon}},\ }\href {\doibase
  https://doi.org/10.1021/acs.langmuir.9b03036} {\bibfield  {journal} {\bibinfo
   {journal} {Langmuir}\ }\textbf {\bibinfo {volume} {36}},\ \bibinfo {pages}
  {6963} (\bibinfo {year} {2020})}\BibitemShut {NoStop}%
\bibitem [{\citenamefont {Lee}\ \emph {et~al.}(2023)\citenamefont {Lee},
  \citenamefont {Thome}, \citenamefont {Cruse}, \citenamefont {Ganguly},
  \citenamefont {Gupta},\ and\ \citenamefont {{Shields,
  IV}}}]{AnkurNanoscale2023}%
  \BibitemOpen
  \bibfield  {author} {\bibinfo {author} {\bibfnamefont {J.~G.}\ \bibnamefont
  {Lee}}, \bibinfo {author} {\bibfnamefont {C.~P.}\ \bibnamefont {Thome}},
  \bibinfo {author} {\bibfnamefont {Z.~A.}\ \bibnamefont {Cruse}}, \bibinfo
  {author} {\bibfnamefont {A.}~\bibnamefont {Ganguly}}, \bibinfo {author}
  {\bibfnamefont {A.}~\bibnamefont {Gupta}}, \ and\ \bibinfo {author}
  {\bibfnamefont {C.~W.}\ \bibnamefont {{Shields, IV}}},\ }\href {\doibase
  https://doi.org/10.1039/D3NR03744D} {\bibfield  {journal} {\bibinfo
  {journal} {Nanoscale}\ }\textbf {\bibinfo {volume} {15}},\ \bibinfo {pages}
  {16268} (\bibinfo {year} {2023})}\BibitemShut {NoStop}%
\bibitem [{\citenamefont {Vogel}\ \emph {et~al.}(2015)\citenamefont {Vogel},
  \citenamefont {Retsch}, \citenamefont {Fustin}, \citenamefont {{del Campo}},\
  and\ \citenamefont {Jonas}}]{selfAssemblyreviewB}%
  \BibitemOpen
  \bibfield  {author} {\bibinfo {author} {\bibfnamefont {N.}~\bibnamefont
  {Vogel}}, \bibinfo {author} {\bibfnamefont {M.}~\bibnamefont {Retsch}},
  \bibinfo {author} {\bibfnamefont {C.}~\bibnamefont {Fustin}}, \bibinfo
  {author} {\bibfnamefont {A.}~\bibnamefont {{del Campo}}}, \ and\ \bibinfo
  {author} {\bibfnamefont {U.}~\bibnamefont {Jonas}},\ }\href {\doibase
  https://doi.org/10.1021/cr400081d} {\bibfield  {journal} {\bibinfo  {journal}
  {Chem. Rev.}\ }\textbf {\bibinfo {volume} {115}},\ \bibinfo {pages} {6265}
  (\bibinfo {year} {2015})}\BibitemShut {NoStop}%
\bibitem [{\citenamefont {Lin}\ \emph {et~al.}(2018)\citenamefont {Lin},
  \citenamefont {Gao}, \citenamefont {Chen}, \citenamefont {Lin},\ and\
  \citenamefont {He}}]{selfAssemblyreviewA}%
  \BibitemOpen
  \bibfield  {author} {\bibinfo {author} {\bibfnamefont {Z.}~\bibnamefont
  {Lin}}, \bibinfo {author} {\bibfnamefont {C.}~\bibnamefont {Gao}}, \bibinfo
  {author} {\bibfnamefont {M.}~\bibnamefont {Chen}}, \bibinfo {author}
  {\bibfnamefont {X.}~\bibnamefont {Lin}}, \ and\ \bibinfo {author}
  {\bibfnamefont {Q.}~\bibnamefont {He}},\ }\href {\doibase
  https://doi.org/10.1016/j.cocis.2018.01.006} {\bibfield  {journal} {\bibinfo
  {journal} {Curr. Opin. Colloid Interface Sci.}\ }\textbf {\bibinfo {volume}
  {35}},\ \bibinfo {pages} {51} (\bibinfo {year} {2018})}\BibitemShut {NoStop}%
\bibitem [{\citenamefont {Corni}\ \emph {et~al.}(2008)\citenamefont {Corni},
  \citenamefont {Ryan},\ and\ \citenamefont {Boccaccini}}]{EPD1}%
  \BibitemOpen
  \bibfield  {author} {\bibinfo {author} {\bibfnamefont {I.}~\bibnamefont
  {Corni}}, \bibinfo {author} {\bibfnamefont {M.~P.}\ \bibnamefont {Ryan}}, \
  and\ \bibinfo {author} {\bibfnamefont {A.~R.}\ \bibnamefont {Boccaccini}},\
  }\href {\doibase https://doi.org/10.1016/j.jeurceramsoc.2007.12.011}
  {\bibfield  {journal} {\bibinfo  {journal} {J. Eur. Ceram. Soc.}\ }\textbf
  {\bibinfo {volume} {28}},\ \bibinfo {pages} {1353} (\bibinfo {year}
  {2008})}\BibitemShut {NoStop}%
\bibitem [{\citenamefont {Hajizadeh}\ \emph {et~al.}(2022)\citenamefont
  {Hajizadeh}, \citenamefont {Shahalizade}, \citenamefont {Riahifar},
  \citenamefont {{Sahba Yaghmaee}}, \citenamefont {Raissi}, \citenamefont
  {Gholam}, \citenamefont {Aghaei}, \citenamefont {Rahimisheikh},\ and\
  \citenamefont {{Sadeghi Ghazvini}}}]{EPD2}%
  \BibitemOpen
  \bibfield  {author} {\bibinfo {author} {\bibfnamefont {A.}~\bibnamefont
  {Hajizadeh}}, \bibinfo {author} {\bibfnamefont {T.}~\bibnamefont
  {Shahalizade}}, \bibinfo {author} {\bibfnamefont {R.}~\bibnamefont
  {Riahifar}}, \bibinfo {author} {\bibfnamefont {M.}~\bibnamefont {{Sahba
  Yaghmaee}}}, \bibinfo {author} {\bibfnamefont {B.}~\bibnamefont {Raissi}},
  \bibinfo {author} {\bibfnamefont {S.}~\bibnamefont {Gholam}}, \bibinfo
  {author} {\bibfnamefont {A.}~\bibnamefont {Aghaei}}, \bibinfo {author}
  {\bibfnamefont {S.}~\bibnamefont {Rahimisheikh}}, \ and\ \bibinfo {author}
  {\bibfnamefont {A.}~\bibnamefont {{Sadeghi Ghazvini}}},\ }\href {\doibase
  https://doi.org/10.1016/j.jpowsour.2022.231448} {\bibfield  {journal}
  {\bibinfo  {journal} {J. Power Sources}\ }\textbf {\bibinfo {volume} {535}},\
  \bibinfo {pages} {231448} (\bibinfo {year} {2022})}\BibitemShut {NoStop}%
\bibitem [{\citenamefont {Liu}\ \emph {et~al.}(2022)\citenamefont {Liu},
  \citenamefont {Li}, \citenamefont {Li}, \citenamefont {Chen}, \citenamefont
  {Zhao}, \citenamefont {Liu}, \citenamefont {{Wei Sun}},\ and\ \citenamefont
  {Wang}}]{EPD3}%
  \BibitemOpen
  \bibfield  {author} {\bibinfo {author} {\bibfnamefont {W.}~\bibnamefont
  {Liu}}, \bibinfo {author} {\bibfnamefont {Y.}~\bibnamefont {Li}}, \bibinfo
  {author} {\bibfnamefont {D.}~\bibnamefont {Li}}, \bibinfo {author}
  {\bibfnamefont {L.}~\bibnamefont {Chen}}, \bibinfo {author} {\bibfnamefont
  {J.}~\bibnamefont {Zhao}}, \bibinfo {author} {\bibfnamefont {P.}~\bibnamefont
  {Liu}}, \bibinfo {author} {\bibfnamefont {X.}~\bibnamefont {{Wei Sun}}}, \
  and\ \bibinfo {author} {\bibfnamefont {G.}~\bibnamefont {Wang}},\ }\href
  {\doibase 10.1002/smll.202107629} {\bibfield  {journal} {\bibinfo  {journal}
  {Small}\ }\textbf {\bibinfo {volume} {18}},\ \bibinfo {pages} {2107629}
  (\bibinfo {year} {2022})}\BibitemShut {NoStop}%
\bibitem [{\citenamefont {Bazant}\ \emph {et~al.}(2009)\citenamefont {Bazant},
  \citenamefont {Kilic}, \citenamefont {Storey},\ and\ \citenamefont
  {Ajdari}}]{Bazantreview2009}%
  \BibitemOpen
  \bibfield  {author} {\bibinfo {author} {\bibfnamefont {M.~Z.}\ \bibnamefont
  {Bazant}}, \bibinfo {author} {\bibfnamefont {M.~S.}\ \bibnamefont {Kilic}},
  \bibinfo {author} {\bibfnamefont {B.~D.}\ \bibnamefont {Storey}}, \ and\
  \bibinfo {author} {\bibfnamefont {A.}~\bibnamefont {Ajdari}},\ }\href
  {\doibase https://doi.org/10.1016/j.cis.2009.10.001} {\bibfield  {journal}
  {\bibinfo  {journal} {Adv. Colloid Interface Sci.}\ }\textbf {\bibinfo
  {volume} {152}},\ \bibinfo {pages} {48} (\bibinfo {year} {2009})}\BibitemShut
  {NoStop}%
\bibitem [{\citenamefont {Squires}(2009)}]{Squires2009}%
  \BibitemOpen
  \bibfield  {author} {\bibinfo {author} {\bibfnamefont {T.~M.}\ \bibnamefont
  {Squires}},\ }\href {\doibase https://doi.org/10.1039/B906909G} {\bibfield
  {journal} {\bibinfo  {journal} {Lab Chip}\ }\textbf {\bibinfo {volume} {9}},\
  \bibinfo {pages} {2477} (\bibinfo {year} {2009})}\BibitemShut {NoStop}%
\bibitem [{\citenamefont {Garcia-Sanchez}\ \emph {et~al.}(2009)\citenamefont
  {Garcia-Sanchez}, \citenamefont {Ramos}, \citenamefont {Gonzalez},
  \citenamefont {Green},\ and\ \citenamefont {Morgan}}]{Garcia-Sanchez2009}%
  \BibitemOpen
  \bibfield  {author} {\bibinfo {author} {\bibfnamefont {P.}~\bibnamefont
  {Garcia-Sanchez}}, \bibinfo {author} {\bibfnamefont {A.}~\bibnamefont
  {Ramos}}, \bibinfo {author} {\bibfnamefont {A.}~\bibnamefont {Gonzalez}},
  \bibinfo {author} {\bibfnamefont {N.~G.}\ \bibnamefont {Green}}, \ and\
  \bibinfo {author} {\bibfnamefont {H.}~\bibnamefont {Morgan}},\ }\href
  {\doibase https://doi.org/10.1021/la803651e} {\bibfield  {journal} {\bibinfo
  {journal} {Langmuir}\ }\textbf {\bibinfo {volume} {25}},\ \bibinfo {pages}
  {4988} (\bibinfo {year} {2009})}\BibitemShut {NoStop}%
\bibitem [{\citenamefont {Hashemi}\ \emph {et~al.}(2020)\citenamefont
  {Hashemi}, \citenamefont {Miller},\ and\ \citenamefont
  {Ristenpart}}]{Aref2020}%
  \BibitemOpen
  \bibfield  {author} {\bibinfo {author} {\bibfnamefont {A.}~\bibnamefont
  {Hashemi}}, \bibinfo {author} {\bibfnamefont {G.~H.}\ \bibnamefont {Miller}},
  \ and\ \bibinfo {author} {\bibfnamefont {W.~D.}\ \bibnamefont {Ristenpart}},\
  }\href {\doibase https://doi.org/10.1103/PhysRevFluids.5.013702} {\bibfield
  {journal} {\bibinfo  {journal} {Phys. Rev. Fluids}\ }\textbf {\bibinfo
  {volume} {5}},\ \bibinfo {pages} {013702} (\bibinfo {year}
  {2020})}\BibitemShut {NoStop}%
\bibitem [{\citenamefont {Woehl}\ \emph {et~al.}(2014)\citenamefont {Woehl},
  \citenamefont {Heatley}, \citenamefont {Dutcher}, \citenamefont {Talken},\
  and\ \citenamefont {Ristenpart}}]{Woehl2014}%
  \BibitemOpen
  \bibfield  {author} {\bibinfo {author} {\bibfnamefont {T.~J.}\ \bibnamefont
  {Woehl}}, \bibinfo {author} {\bibfnamefont {K.~L.}\ \bibnamefont {Heatley}},
  \bibinfo {author} {\bibfnamefont {C.~S.}\ \bibnamefont {Dutcher}}, \bibinfo
  {author} {\bibfnamefont {N.~H.}\ \bibnamefont {Talken}}, \ and\ \bibinfo
  {author} {\bibfnamefont {W.~D.}\ \bibnamefont {Ristenpart}},\ }\href
  {\doibase https://doi.org/10.1021/la4048243} {\bibfield  {journal} {\bibinfo
  {journal} {Langmuir}\ }\textbf {\bibinfo {volume} {30}},\ \bibinfo {pages}
  {4887} (\bibinfo {year} {2014})}\BibitemShut {NoStop}%
\bibitem [{\citenamefont {Hashemi}\ \emph {et~al.}(2018)\citenamefont
  {Hashemi}, \citenamefont {Bukosky}, \citenamefont {Rader}, \citenamefont
  {Ristenpart},\ and\ \citenamefont {Miller}}]{Aref2018}%
  \BibitemOpen
  \bibfield  {author} {\bibinfo {author} {\bibfnamefont {A.}~\bibnamefont
  {Hashemi}}, \bibinfo {author} {\bibfnamefont {S.~C.}\ \bibnamefont
  {Bukosky}}, \bibinfo {author} {\bibfnamefont {S.~P.}\ \bibnamefont {Rader}},
  \bibinfo {author} {\bibfnamefont {W.~D.}\ \bibnamefont {Ristenpart}}, \ and\
  \bibinfo {author} {\bibfnamefont {G.~H.}\ \bibnamefont {Miller}},\ }\href
  {\doibase https://doi.org/10.1103/PhysRevLett.121.185504} {\bibfield
  {journal} {\bibinfo  {journal} {Phys. Rev. Lett.}\ }\textbf {\bibinfo
  {volume} {121}},\ \bibinfo {pages} {185504} (\bibinfo {year}
  {2018})}\BibitemShut {NoStop}%
\bibitem [{\citenamefont {Woehl}\ \emph {et~al.}(2015)\citenamefont {Woehl},
  \citenamefont {Chen}, \citenamefont {Heatley}, \citenamefont {Talken},
  \citenamefont {Bukosky}, \citenamefont {Dutcher},\ and\ \citenamefont
  {Ristenpart}}]{Woehl2015}%
  \BibitemOpen
  \bibfield  {author} {\bibinfo {author} {\bibfnamefont {T.~J.}\ \bibnamefont
  {Woehl}}, \bibinfo {author} {\bibfnamefont {B.~J.}\ \bibnamefont {Chen}},
  \bibinfo {author} {\bibfnamefont {K.~L.}\ \bibnamefont {Heatley}}, \bibinfo
  {author} {\bibfnamefont {N.~H.}\ \bibnamefont {Talken}}, \bibinfo {author}
  {\bibfnamefont {S.~C.}\ \bibnamefont {Bukosky}}, \bibinfo {author}
  {\bibfnamefont {C.~S.}\ \bibnamefont {Dutcher}}, \ and\ \bibinfo {author}
  {\bibfnamefont {W.~D.}\ \bibnamefont {Ristenpart}},\ }\href {\doibase
  https://doi.org/10.1103/PhysRevX.5.011023} {\bibfield  {journal} {\bibinfo
  {journal} {Phys. Rev. X}\ }\textbf {\bibinfo {volume} {5}},\ \bibinfo {pages}
  {011023} (\bibinfo {year} {2015})}\BibitemShut {NoStop}%
\bibitem [{\citenamefont {Bukosky}\ and\ \citenamefont
  {Ristenpart}(2015)}]{Bukosky2015}%
  \BibitemOpen
  \bibfield  {author} {\bibinfo {author} {\bibfnamefont {S.~C.}\ \bibnamefont
  {Bukosky}}\ and\ \bibinfo {author} {\bibfnamefont {W.~D.}\ \bibnamefont
  {Ristenpart}},\ }\href {\doibase
  https://doi.org/10.1021/acs.langmuir.5b02432} {\bibfield  {journal} {\bibinfo
   {journal} {Langmuir}\ }\textbf {\bibinfo {volume} {31}},\ \bibinfo {pages}
  {9742} (\bibinfo {year} {2015})}\BibitemShut {NoStop}%
\bibitem [{\citenamefont {Bukosky}\ \emph {et~al.}(2019)\citenamefont
  {Bukosky}, \citenamefont {Hashemi}, \citenamefont {Rader}, \citenamefont
  {Mora}, \citenamefont {Miller},\ and\ \citenamefont
  {Ristenpart}}]{Bukosky2019}%
  \BibitemOpen
  \bibfield  {author} {\bibinfo {author} {\bibfnamefont {S.~C.}\ \bibnamefont
  {Bukosky}}, \bibinfo {author} {\bibfnamefont {A.}~\bibnamefont {Hashemi}},
  \bibinfo {author} {\bibfnamefont {S.~P.}\ \bibnamefont {Rader}}, \bibinfo
  {author} {\bibfnamefont {J.}~\bibnamefont {Mora}}, \bibinfo {author}
  {\bibfnamefont {G.~H.}\ \bibnamefont {Miller}}, \ and\ \bibinfo {author}
  {\bibfnamefont {W.~D.}\ \bibnamefont {Ristenpart}},\ }\href {\doibase
  https://doi.org/10.1021/acs.langmuir.9b00313} {\bibfield  {journal} {\bibinfo
   {journal} {Langmuir}\ }\textbf {\bibinfo {volume} {35}},\ \bibinfo {pages}
  {6971} (\bibinfo {year} {2019})}\BibitemShut {NoStop}%
\bibitem [{\citenamefont {{Chandan Naru}}(2021)}]{CompTimeWaterOil}%
  \BibitemOpen
  \bibfield  {author} {\bibinfo {author} {\bibfnamefont {M.}~\bibnamefont
  {{Chandan Naru}}},\ }\href@noop {} {\emph {\bibinfo {title} {Numerical
  simulation of a water/oil emulsion in a multiscale/multiphysics context}}}\
  (\bibinfo  {publisher} {Sorbonne Université},\ \bibinfo {year}
  {2021})\BibitemShut {NoStop}%
\bibitem [{\citenamefont {Liu}\ \emph {et~al.}(2021)\citenamefont {Liu},
  \citenamefont {Ng}, \citenamefont {Chong}, \citenamefont {Lohse},\ and\
  \citenamefont {Verzicco}}]{CompTimeTurbulent}%
  \BibitemOpen
  \bibfield  {author} {\bibinfo {author} {\bibfnamefont {H.-R.}\ \bibnamefont
  {Liu}}, \bibinfo {author} {\bibfnamefont {C.~S.}\ \bibnamefont {Ng}},
  \bibinfo {author} {\bibfnamefont {K.~L.}\ \bibnamefont {Chong}}, \bibinfo
  {author} {\bibfnamefont {D.}~\bibnamefont {Lohse}}, \ and\ \bibinfo {author}
  {\bibfnamefont {R.}~\bibnamefont {Verzicco}},\ }\href {\doibase
  https://doi.org/10.1016/j.jcp.2021.110659} {\bibfield  {journal} {\bibinfo
  {journal} {J. Comput. Phys.}\ }\textbf {\bibinfo {volume} {446}},\ \bibinfo
  {pages} {110659} (\bibinfo {year} {2021})}\BibitemShut {NoStop}%
\bibitem [{\citenamefont {Geitani}\ \emph {et~al.}(2023)\citenamefont
  {Geitani}, \citenamefont {Golshan},\ and\ \citenamefont
  {Blais}}]{FluidizedBedBruno}%
  \BibitemOpen
  \bibfield  {author} {\bibinfo {author} {\bibfnamefont {T.~E.}\ \bibnamefont
  {Geitani}}, \bibinfo {author} {\bibfnamefont {S.}~\bibnamefont {Golshan}}, \
  and\ \bibinfo {author} {\bibfnamefont {B.}~\bibnamefont {Blais}},\ }\href
  {\doibase https://doi.org/10.1021/acs.iecr.2c03546} {\bibfield  {journal}
  {\bibinfo  {journal} {Ind. Eng. Chem. Res.}\ }\textbf {\bibinfo {volume}
  {62}},\ \bibinfo {pages} {1141} (\bibinfo {year} {2023})}\BibitemShut
  {NoStop}%
\bibitem [{\citenamefont {Naghdi}\ \emph {et~al.}(2024)\citenamefont {Naghdi},
  \citenamefont {Arabi},\ and\ \citenamefont {{Molaei
  Dehkordi}}}]{FluidizedBedMolaeia}%
  \BibitemOpen
  \bibfield  {author} {\bibinfo {author} {\bibfnamefont {H.}~\bibnamefont
  {Naghdi}}, \bibinfo {author} {\bibfnamefont {Y.}~\bibnamefont {Arabi}}, \
  and\ \bibinfo {author} {\bibfnamefont {A.}~\bibnamefont {{Molaei
  Dehkordi}}},\ }\href {\doibase https://doi.org/10.1021/acs.iecr.3c03818}
  {\bibfield  {journal} {\bibinfo  {journal} {Ind. Eng. Chem. Res.}\ }\textbf
  {\bibinfo {volume} {63}},\ \bibinfo {pages} {1690} (\bibinfo {year}
  {2024})}\BibitemShut {NoStop}%
\bibitem [{\citenamefont {Sarafan}\ and\ \citenamefont {{Molaei
  Dehkordi}}(2023)}]{FluidizedBedMolaeib}%
  \BibitemOpen
  \bibfield  {author} {\bibinfo {author} {\bibfnamefont {K.}~\bibnamefont
  {Sarafan}}\ and\ \bibinfo {author} {\bibfnamefont {A.}~\bibnamefont {{Molaei
  Dehkordi}}},\ }\href {\doibase https://doi.org/10.1021/acs.iecr.3c02925}
  {\bibfield  {journal} {\bibinfo  {journal} {Ind. Eng. Chem. Res.}\ }\textbf
  {\bibinfo {volume} {62}},\ \bibinfo {pages} {22115} (\bibinfo {year}
  {2023})}\BibitemShut {NoStop}%
\bibitem [{\citenamefont {Kochkov}\ \emph {et~al.}(2021)\citenamefont
  {Kochkov}, \citenamefont {Smith}, \citenamefont {Alieva}, \citenamefont
  {Wang}, \citenamefont {Brenner},\ and\ \citenamefont {Hoyer}}]{Kochkov2021}%
  \BibitemOpen
  \bibfield  {author} {\bibinfo {author} {\bibfnamefont {D.}~\bibnamefont
  {Kochkov}}, \bibinfo {author} {\bibfnamefont {J.~A.}\ \bibnamefont {Smith}},
  \bibinfo {author} {\bibfnamefont {A.}~\bibnamefont {Alieva}}, \bibinfo
  {author} {\bibfnamefont {Q.}~\bibnamefont {Wang}}, \bibinfo {author}
  {\bibfnamefont {M.~P.}\ \bibnamefont {Brenner}}, \ and\ \bibinfo {author}
  {\bibfnamefont {S.}~\bibnamefont {Hoyer}},\ }\href {\doibase
  https://doi.org/10.1073/pnas.2101784118} {\bibfield  {journal} {\bibinfo
  {journal} {PNAS}\ }\textbf {\bibinfo {volume} {118}},\ \bibinfo {pages}
  {e2101784118} (\bibinfo {year} {2021})}\BibitemShut {NoStop}%
\bibitem [{\citenamefont {Sanchez-Gonzalez}\ \emph {et~al.}(2020)\citenamefont
  {Sanchez-Gonzalez}, \citenamefont {Godwin}, \citenamefont {Pfaff},
  \citenamefont {Ying}, \citenamefont {Leskovec},\ and\ \citenamefont
  {Battaglia}}]{GNS-deepmind}%
  \BibitemOpen
  \bibfield  {author} {\bibinfo {author} {\bibfnamefont {A.}~\bibnamefont
  {Sanchez-Gonzalez}}, \bibinfo {author} {\bibfnamefont {J.}~\bibnamefont
  {Godwin}}, \bibinfo {author} {\bibfnamefont {T.}~\bibnamefont {Pfaff}},
  \bibinfo {author} {\bibfnamefont {R.}~\bibnamefont {Ying}}, \bibinfo {author}
  {\bibfnamefont {J.}~\bibnamefont {Leskovec}}, \ and\ \bibinfo {author}
  {\bibfnamefont {P.~W.}\ \bibnamefont {Battaglia}},\ }\href
  {https://proceedings.mlr.press/v119/sanchez-gonzalez20a/sanchez-gonzalez20a.pdf}
  {\bibfield  {journal} {\bibinfo  {journal} {Proceedings of the
  37\textsuperscript{th} International Conference on Machine Learning, PMLR}\
  }\textbf {\bibinfo {volume} {119}},\ \bibinfo {pages} {8459} (\bibinfo {year}
  {2020})}\BibitemShut {NoStop}%
\bibitem [{\citenamefont {Gilmer}\ \emph {et~al.}(2017)\citenamefont {Gilmer},
  \citenamefont {Schoenholz}, \citenamefont {Riley}, \citenamefont {Vinyals},\
  and\ \citenamefont {Dahl}}]{GNN-neuralmp}%
  \BibitemOpen
  \bibfield  {author} {\bibinfo {author} {\bibfnamefont {J.}~\bibnamefont
  {Gilmer}}, \bibinfo {author} {\bibfnamefont {S.~S.}\ \bibnamefont
  {Schoenholz}}, \bibinfo {author} {\bibfnamefont {P.~F.}\ \bibnamefont
  {Riley}}, \bibinfo {author} {\bibfnamefont {O.}~\bibnamefont {Vinyals}}, \
  and\ \bibinfo {author} {\bibfnamefont {G.~E.}\ \bibnamefont {Dahl}},\ }\href
  {http://proceedings.mlr.press/v70/gilmer17a/gilmer17a.pdf} {\bibfield
  {journal} {\bibinfo  {journal} {Proceedings of the 34\textsuperscript{th}
  International Conference on Machine Learning, PMLR}\ }\textbf {\bibinfo
  {volume} {70}},\ \bibinfo {pages} {1263} (\bibinfo {year}
  {2017})}\BibitemShut {NoStop}%
\bibitem [{\citenamefont {Monaghan}(1992)}]{SPH}%
  \BibitemOpen
  \bibfield  {author} {\bibinfo {author} {\bibfnamefont {J.~J.}\ \bibnamefont
  {Monaghan}},\ }\href {https://doi.org/10.1146/annurev.aa.30.090192.002551}
  {\bibfield  {journal} {\bibinfo  {journal} {Annual Review of Astronomy \&
  Astrophysics}\ }\textbf {\bibinfo {volume} {30}},\ \bibinfo {pages} {543}
  (\bibinfo {year} {1992})}\BibitemShut {NoStop}%
\bibitem [{\citenamefont {Muller}\ \emph {et~al.}(2007)\citenamefont {Muller},
  \citenamefont {Heidelberger}, \citenamefont {Hennix},\ and\ \citenamefont
  {Ratcliff}}]{PBD}%
  \BibitemOpen
  \bibfield  {author} {\bibinfo {author} {\bibfnamefont {M.}~\bibnamefont
  {Muller}}, \bibinfo {author} {\bibfnamefont {B.}~\bibnamefont
  {Heidelberger}}, \bibinfo {author} {\bibfnamefont {M.}~\bibnamefont
  {Hennix}}, \ and\ \bibinfo {author} {\bibfnamefont {J.}~\bibnamefont
  {Ratcliff}},\ }\href {https://doi.org/10.1016/j.jvcir.2007.01.005} {\bibfield
   {journal} {\bibinfo  {journal} {Journal of Visual Communication \& Image
  Representation}\ }\textbf {\bibinfo {volume} {18}},\ \bibinfo {pages} {109}
  (\bibinfo {year} {2007})}\BibitemShut {NoStop}%
\bibitem [{\citenamefont {Sulsky}\ \emph {et~al.}(1995)\citenamefont {Sulsky},
  \citenamefont {Zhou},\ and\ \citenamefont {Schreyer}}]{MPM}%
  \BibitemOpen
  \bibfield  {author} {\bibinfo {author} {\bibfnamefont {D.}~\bibnamefont
  {Sulsky}}, \bibinfo {author} {\bibfnamefont {S.}~\bibnamefont {Zhou}}, \ and\
  \bibinfo {author} {\bibfnamefont {H.~L.}\ \bibnamefont {Schreyer}},\ }\href
  {https://doi.org/10.1016/0010-4655(94)00170-7} {\bibfield  {journal}
  {\bibinfo  {journal} {Computer Physics Communications}\ }\textbf {\bibinfo
  {volume} {87}},\ \bibinfo {pages} {236} (\bibinfo {year} {1995})}\BibitemShut
  {NoStop}%
\bibitem [{\citenamefont {Sergeev}\ \emph {et~al.}(2024)\citenamefont
  {Sergeev}, \citenamefont {Kiselev},\ and\ \citenamefont
  {Makarov}}]{Sergeev2024}%
  \BibitemOpen
  \bibfield  {author} {\bibinfo {author} {\bibfnamefont {P.}~\bibnamefont
  {Sergeev}}, \bibinfo {author} {\bibfnamefont {D.}~\bibnamefont {Kiselev}}, \
  and\ \bibinfo {author} {\bibfnamefont {I.}~\bibnamefont {Makarov}},\ }in\
  \href {\doibase https://doi.org/10.1109/INES63318.2024.10629094} {\emph
  {\bibinfo {booktitle} {28th IEEE International Conference on Intelligent
  Engineering Systems (INES)}}}\ (\bibinfo  {publisher} {IEEE},\ \bibinfo
  {year} {2024})\BibitemShut {NoStop}%
\bibitem [{\citenamefont {Bhattoo}\ \emph {et~al.}(2023)\citenamefont
  {Bhattoo}, \citenamefont {Ranu},\ and\ \citenamefont
  {Krishnan}}]{Bhattoo2023}%
  \BibitemOpen
  \bibfield  {author} {\bibinfo {author} {\bibfnamefont {R.}~\bibnamefont
  {Bhattoo}}, \bibinfo {author} {\bibfnamefont {S.}~\bibnamefont {Ranu}}, \
  and\ \bibinfo {author} {\bibfnamefont {N.~M.~A.}\ \bibnamefont {Krishnan}},\
  }\href {\doibase https://doi.org/10.1088/2632-2153/acb03e} {\bibfield
  {journal} {\bibinfo  {journal} {Mach. Learn.: Sci. Technol.}\ }\textbf
  {\bibinfo {volume} {4}},\ \bibinfo {pages} {015003} (\bibinfo {year}
  {2023})}\BibitemShut {NoStop}%
\bibitem [{\citenamefont {Ma}\ \emph {et~al.}(2022)\citenamefont {Ma},
  \citenamefont {Ye},\ and\ \citenamefont {Pan}}]{Ma2022}%
  \BibitemOpen
  \bibfield  {author} {\bibinfo {author} {\bibfnamefont {Z.}~\bibnamefont
  {Ma}}, \bibinfo {author} {\bibfnamefont {Z.}~\bibnamefont {Ye}}, \ and\
  \bibinfo {author} {\bibfnamefont {W.}~\bibnamefont {Pan}},\ }\href {\doibase
  https://doi.org/10.1016/j.cma.2022.115496} {\bibfield  {journal} {\bibinfo
  {journal} {Comput. Methods Appl. Mech. Engrg.}\ }\textbf {\bibinfo {volume}
  {400}},\ \bibinfo {pages} {115496} (\bibinfo {year} {2022})}\BibitemShut
  {NoStop}%
\bibitem [{\citenamefont {Martinkus}\ \emph {et~al.}(2021)\citenamefont
  {Martinkus}, \citenamefont {Lucchi},\ and\ \citenamefont
  {Perraudin}}]{Martinkus2021}%
  \BibitemOpen
  \bibfield  {author} {\bibinfo {author} {\bibfnamefont {K.}~\bibnamefont
  {Martinkus}}, \bibinfo {author} {\bibfnamefont {A.}~\bibnamefont {Lucchi}}, \
  and\ \bibinfo {author} {\bibfnamefont {N.}~\bibnamefont {Perraudin}},\ }in\
  \href {\doibase
  file:///home/ahashem2/Downloads/17078-Article%20Text-20572-1-2-20210518-1.pdf}
  {\emph {\bibinfo {booktitle} {The Thirty-Fifth AAAI Conference on Artificial
  Intelligence}}}\ (\bibinfo  {publisher} {AAAI},\ \bibinfo {year}
  {2021})\BibitemShut {NoStop}%
\bibitem [{\citenamefont {Filiatraut}\ \emph {et~al.}(2023)\citenamefont
  {Filiatraut}, \citenamefont {Mianroodi}, \citenamefont {Siboni},\ and\
  \citenamefont {Zanjani}}]{Filiatraut2023}%
  \BibitemOpen
  \bibfield  {author} {\bibinfo {author} {\bibfnamefont {A.~N.}\ \bibnamefont
  {Filiatraut}}, \bibinfo {author} {\bibfnamefont {J.~R.}\ \bibnamefont
  {Mianroodi}}, \bibinfo {author} {\bibfnamefont {N.~H.}\ \bibnamefont
  {Siboni}}, \ and\ \bibinfo {author} {\bibfnamefont {M.~B.}\ \bibnamefont
  {Zanjani}},\ }\href {\doibase https://doi.org/10.1063/5.0175062} {\bibfield
  {journal} {\bibinfo  {journal} {J. Appl. Phys.}\ }\textbf {\bibinfo {volume}
  {134}},\ \bibinfo {pages} {234702} (\bibinfo {year} {2023})}\BibitemShut
  {NoStop}%
\bibitem [{\citenamefont {Kumar}\ and\ \citenamefont
  {Vantassel}(2023)}]{GNS-pytorch}%
  \BibitemOpen
  \bibfield  {author} {\bibinfo {author} {\bibfnamefont {K.}~\bibnamefont
  {Kumar}}\ and\ \bibinfo {author} {\bibfnamefont {J.}~\bibnamefont
  {Vantassel}},\ }\href {https://doi.org/10.21105/joss.05025} {\bibfield
  {journal} {\bibinfo  {journal} {Journal of Open Source Software}\ }\textbf
  {\bibinfo {volume} {8}},\ \bibinfo {pages} {5025} (\bibinfo {year}
  {2023})}\BibitemShut {NoStop}%
\bibitem [{\citenamefont {Ruiz-Garcia}\ \emph {et~al.}(2024)\citenamefont
  {Ruiz-Garcia}, \citenamefont {{Barriuso G.}}, \citenamefont {Alexander},
  \citenamefont {Aarts}, \citenamefont {Ghiringhelli},\ and\ \citenamefont
  {Valeriani}}]{ActiveNet}%
  \BibitemOpen
  \bibfield  {author} {\bibinfo {author} {\bibfnamefont {M.}~\bibnamefont
  {Ruiz-Garcia}}, \bibinfo {author} {\bibfnamefont {C.~M.}\ \bibnamefont
  {{Barriuso G.}}}, \bibinfo {author} {\bibfnamefont {L.~C.}\ \bibnamefont
  {Alexander}}, \bibinfo {author} {\bibfnamefont {D.~G. A.~L.}\ \bibnamefont
  {Aarts}}, \bibinfo {author} {\bibfnamefont {L.~M.}\ \bibnamefont
  {Ghiringhelli}}, \ and\ \bibinfo {author} {\bibfnamefont {C.}~\bibnamefont
  {Valeriani}},\ }\href {\doibase https://doi.org/10.1103/PhysRevE.109.064611}
  {\bibfield  {journal} {\bibinfo  {journal} {Phys. Rev. E}\ }\textbf {\bibinfo
  {volume} {109}},\ \bibinfo {pages} {064611} (\bibinfo {year}
  {2024})}\BibitemShut {NoStop}%
\end{thebibliography}

%

\end{document}